\documentclass[aps,prd,superscript adress,showpacs,twocolumn]{revtex4-1}
\usepackage{graphicx}
\usepackage[caption=false]{subfig}
\usepackage{epstopdf}
\usepackage{amsmath}
\usepackage{amsfonts} 
\usepackage{amssymb}
\usepackage{latexsym}
\usepackage{hyperref}
\usepackage[english]{babel}
\usepackage[utf8]{inputenc}
\usepackage[colorinlistoftodos]{todonotes}
\usepackage{xcolor}
\usepackage{slashed}
\usepackage{feynmp}

\usepackage{bbold}
\usepackage{eufrak}

\begin{document}
\title{Supersymmetry Breaking at Finite Temperature in a Susy Harmonic Oscillator with Interaction.}

\author{$^1$$^,$$^2$C. Marques}
\author{$^1$G S Dias}
\author{$^3$F C Khanna}
\author{$^4$H. S. Chavez}
\affiliation{$^1$Department of Physics - IFES - Instituto Federal do Esp\'{i}rito Santo (IFES), \\ Av. Vit\'{o}ria 1729, Jucutuquara, Vit\'{o}ria, ES, Brazil, CEP 29040-780}
\affiliation{$^2$Centro Brasileiro de Pesquisas F\'{i}sicas (CBPF) Rua Dr. Xavier \\ Sigaud 150, Urca, RJ, Brazil, CEP 22290-180} 
\affiliation{$^3$Department of Physics - University of Alberta - Edmonton,\\ Alberta, Canada} 
\affiliation{$^4$UCL-Faculdade do Centro Leste, Rodovia ES 010, Km 6 \\ BR 101, Serra-ES, Brazil}

\email{celiom@ifes.edu.br}
\email{gilmar@ifes.edu.br}

\begin{abstract}
Supersymmetry breaking of a supersymmetric harmonic oscillator with polynomial interaction is analyzed. Some thermal effects are studied in the context of Thermo Field Dynamics (TFD). The restored supersymmetry results in non vanishing energy at finite temperatures due the additivity of the thermal effects, while at $T=0$ the energy is zero.
\end{abstract}

\maketitle


\section{Introduction} \label{sec_intro}

Supersymmetry has different characteristics than other internal
symmetries. One of these characteristics is that in contrast with the
internal symmetries, supersymmetry can be broken for theories with
finite fermionic and bosonic degrees of freedom \cite{wquebra,susyquebramecanicaquantica}. Therefore we can think of spontaneous
symmetry breaking in supersymmetric quantum mechanics. Some clear motivations and significant progress about the mechanisms of SUSY breaking in quantum field theory could be seen at \cite{faizal,sud}\cite{sud2}. The interest in
supersymmetry breaking in quantum mechanics \cite{Gil1}-\cite{Grisaru}
becomes obvious, when we verify that some results from the breaking of
supersymmetric quantum mechanics can be generalized and applied to quantum
field theory.
 In the limit of low energy, the underlying field theory
should approach a Galilean invariant supersymmetric field theory \cite{Gil1,Gil2} and, by the Bergmann superselection rules \cite{Bergmann}, such
a field theory would be equivalent to a supersymmetric Schr\"odinger equation
in each particle number sector of the theory.

Supersymmetry at finite temperature has been studied in refs \cite{Girardello}, 
\cite{Van Hove}, \cite{Das}, \cite{Teshima}, \cite{Umezawa} and \cite{Umezawa H}. Nevertheless, the issue of whether supersymmetry is broken at
finite temperature has raised some controversy. In \cite{Girardello} it is
argued that supersymmetry (SUSY) is broken at positive temperature even when
unbroken at $T=0$. Regarding this it has been  suggested \cite{Van Hove}, that when
a change of an operator under SUSY transformation at finite temperature is
considered, one should take into account the Klein operator. When this
operator is incorporated, the author of ref.\cite{Van Hove} shows that the thermal average of
the operator variation is zero for all $T$, thereby maintaining
supersymmetry at finite temperature. On the other hand, considering
this issue within the context of Thermo Field Dynamics
(TFD)  \cite{Umezawa,Takahashi} the SUSY is broken at
finite temperature.  The vacuum expectation
value of the SUSY Hamiltonian at $T = 0$, in the thermal vacuum $|\theta(\beta)\rangle$ (where
$\beta = 1/kT$, $k$ being the Boltzmann constant)  is non-zero at finite temperature. By evaluating the statistical
average of the SUSY Hamiltonian at $T=0$ as its vacuum expectation value in
the thermal vacuum $|0(\beta )\rangle$ (where $\beta =1/kT$, $k$ being the
Boltzmann constant) and showing that it is non-zero at finite temperature.

In this paper following TFD, in $T=0$ we show that for some conditions the
supersymmetric harmonic oscillator is broken getting a bosonic and fermionic
harmonic oscillator, with $\omega _{1}$ and $\omega _{2}$ respectively the
bosonic and fermionic frequencies, and the oposite way is possible. Later we
show that in agreement with the former results, the supersymmetry is broken
at finite temperatures.

Many quantum systems can be considered as an application of this work.  A direct application is in the Searching Neutrino-Nucleus interaction \cite{Neutrino}. For example the two level system of neutron valence vibrations considered in the Coherent Neutrino Nucleus Scattering.

\section{Susy Breaking of the Free Susy Harmonic Oscillator at Finite
Temperature.}

Consider a Hamiltonian \cite{wquebra} in supersymmetric quantum mechanics in
the component form 
\begin{equation}
H=\frac{1}{2}[p^{2}+W^{2}(x)+\sigma _{3}W^{\prime }(x)].  \label{whamil}
\end{equation}

The boson-boson interaction is represented by the term $W^{2}(x)$, and the
boson fermion is represented by $\sigma _{3}W^{\prime }(x)$. Both are
determined by the same function $W(x)$. This property is found in all
supersymmetric models.

As shown in \cite{wquebra}, $W(x)$ varies from the superpotential $V(x)$,
where $W(x)=V^{\prime }(x)$and $V^{\prime }(x)$ is the derivative of $V(x)$.
Sometimes $W(x)$ is also called the superpotential.

The double degeneration that occurs in all levels of energy with $E>0$,
follows directly from the supersymmetric algebra; it does not depend on 
$W(x)$. In addition we find that the energy of the ground state is
non-negative. Similarly this follows from the supersymmetric algebra and it is
independent of the function $W(x)$.

Considering any internal symmetry $S$ with generator $G_{S}$. This symmetry $S$ is exact and is not spontaneously broken if $[H,G_{S}]=0$ and the ground
state $|0\rangle$ is invariant, $G_{S}|0\rangle=0$. In the case of supersymmetry, an
example with exact supersymmetry is the\ free supersymmetric harmonic
oscillator, where $W=\omega _{1}x$ \cite{Ascarraga Osc Harm 1, Ragh Osc Harm
2}. 

In terms of the creation and annihilation operators the Hamiltonian
becomes,
\begin{equation}
H=\omega _{1}(a^{\dagger }a+b^{\dagger }b)
\label{OHarmonicoSusyOpCriacao e Destruicao}
\end{equation}%
where $a^{\dagger }$ and $b^{\dagger }$ are respectively the bosonic and
fermionic creation operators and the dual $a$ and $b$, are respectively the
annihilation operators. The algebra is:

\begin{equation*}
a^{\dagger }\wedge a=-1/2;\text{ \ \ \ }b^{\dagger }\bullet b=1/2
\end{equation*}%
where, $\wedge $ and $\bullet $ are related the commutator and
anticommutador respectively. Following the algorithm of TFD, we double the
Hilbert Space \cite{Khanna} writing, 
\begin{equation}
\hat{H}=\omega _{1}(a^{\dagger }a+b^{\dagger }b)-\omega _{1}(\widetilde{a}%
^{\dagger }\widetilde{a}+\widetilde{b}^{\dagger }\widetilde{b})
\label{OHarmonicoSusyOpCriacao e Destruicao termico}
\end{equation}%
Then we introduce a thermal vacuum so that the statistical average of any
operator is given in the thermal vacuum. Then all Feynman techniques of zero
temperature field theory can be used.

The thermal energy given in \cite{Das}, shows that the thermal vacuum 
$|0(\beta )\rangle$ of the free supersymmetric harmonic oscillator has
non-vanishing energy for a positive temperature. In addition the
interaction will not lead to null energy, therefore supersymmetry will not be restored.

\section{Supersymmetric Oscillator with Interaction}

Consider a general Hamiltonian defined by: 
\begin{equation}
H=F(a^{\dagger },b,b^{\dagger })a+G(a^{\dagger },a,b^{\dagger })b
\label{General model}
\end{equation}%
where $a^{\dagger },a,$ are bosonic fields and $b,b^{\dagger},$ are fermionic
fields, and $F$ is a bosonic polynomial of $a^{\dagger },b,b^{\dagger}$ and 
$G$ is a fermionic polynomial of $a^{\dagger },a,b^{\dagger}$. The following
transformation preserves the structure of the Hamiltonian,

\begin{eqnarray}
a_{2} &=&a+\beta _{1}b^{\dagger}b;\text{\ \ \ \ \ \ \ \ \ \ \ \ \ \ \ }a^{\dagger }_{2}=a^{\dagger }%
+\beta _{1}b^{\dagger}b;  \label{extruture transformation} \\
b_{2} &=&(\exp [\beta _{2}(a^{\dagger }-a)])b;\text{ \ \ \ }b^{\dagger}_{2}=%
b^{\dagger}(\exp [\beta _{2}(a-a^{\dagger})]);\text{\ \ \ }
\end{eqnarray}%
where $\beta_{1}$ and $\beta_{2}$ are real parameters. This transformation
defines the Bogoliubov transformation \cite{wall}.

Restricting the model, eq.(\ref{General model}), in order to consider a
oscillator model with interactions, we define the polynomials $F=\omega _{1}%
a^{\dagger }$ and $G=\omega _{2}b^{\dagger}=\alpha _{1}a^{\dagger }b^{\dagger}%
-\alpha _{2}ab^{\dagger}$ where $\omega _{1},\omega _{2},\alpha _{1},\alpha
_{2}$ are real parameters. The definition of perpendicularity and parallelism in the extended Fock space are: 
\begin{eqnarray}
a^{\dagger }\wedge a=-1/2;\text{\ \ \ \ \ \ }
a^{\dagger }\bullet a=n_{b}+1/2 \text{\ } \\ \nonumber
b^{\dagger}\bullet b=1/2 ; \text{\ \ \ \ \ \ \ \ \ } \nonumber
b^{\dagger}\wedge b=n_{f}-1/2  \text{\ }
\label{susy algebra dos campos componentes}
\end{eqnarray}

$n_{b}$ $\in \mathbb{N}$ and $n_{f}\in \{0,1\}$. This algebra is invariant by duality.

It is possible to establish some conditions over the oscillator 
$H=F(a^{\dagger },b,b^{\dagger })a+G(a^{\dagger },a,b^{\dagger })b$ , in the way that it will be a supersymmetric oscillator. To accomplish this we define $\alpha_{1}=\alpha _{2}$; $\omega _{2}=\frac{(\omega _{1})^{2}+(\alpha _{2})^{2}}{
\omega _{1}}$, turning $H$ a supersymmetric oscillator, that will produce the supercharges and the supersymmetric transformations.

\subsection{Supercharge and Supersymmetric Transformations}

Supercharge follows from the condition $[H,G_{S}]=0.$ The supersymmetric oscillator with interactions Eq.(\ref{General model})
has the following supercharges $G_{S}=a^{\dagger }b\exp \frac{1}{\omega _{1}}
(\alpha _{2}a^{\dagger }-\alpha _{2}a)$ and $G_{S}^{\dagger }=[\exp -\frac{1}{\omega _{1}} (\alpha _{2}a^{\dagger }-\alpha _{2}a)]b^{\dagger }a$.\\
The supersymmetric transformations of the component fields are through $G_{S}$:
\begin{eqnarray*}
\delta _{susy}a =([\exp \frac{1}{\omega _{1}}(\alpha _{2}a^{\dagger
}-\alpha _{2}a)]b \text{\ \ \ \ \ \ \ \ \ \ \ \ \ \ \ \ \ } \\ +\frac{1}{\omega _{1}}\alpha _{2}a^{\dagger }[\exp \frac{1}{%
\omega _{1}}(\alpha _{2}a^{\dagger }-\alpha _{2}a)]b)\epsilon ;\text{\ \ } \\
\delta _{susy}b =0; \text{\ \ \ \ \ \ \ \ \ \ \ \ \ \ \ \ \ \ \ \ \ \ \ \ \ \ \ \ \ \ \ \ \ \ \ \ \ \ \ \ \ \ \ \ \ } \\
\delta _{susy}a^{\dagger } =\frac{1}{\omega _{1}}\alpha _{2}a^{\dagger
}b[\exp \frac{1}{\omega _{1}}(\alpha _{2}a^{\dagger }-\alpha _{2}a)]\epsilon; \text{\ \ \ \ \ } \\
\delta _{susy}b^{\dagger } =\frac{1}{\omega _{1}}a^{\dagger }[\exp \frac{1%
}{\omega _{1}}(\alpha _{2}a^{\dagger }-\alpha _{2}a)]\epsilon ; \text{\ \ \ \ \ \ \ \ \ \ }
\end{eqnarray*}%

where $\epsilon$ is a Grassmann parameter. The conjugation is direct from above.

The interactions terms follow from the polynomial $G(a^{\dagger},a,b^{\dagger })$ giving $H=H_{0}+H_{\text{{\small int}}}$ where

\begin{equation*}
H_{0}=\omega _{1}a^{\dagger }a+\omega _{2}b^{\dagger }b,
\end{equation*}%
and%
\begin{equation}
H_{\text{{\small int}}}=\alpha _{2}a^{\dagger }b^{\dagger }b-\alpha_{2}ab^{\dagger }b.  \label{Hamilt de interacao}
\end{equation}

Although $H$ could be a supersymmetric oscillator, $H_{0}$ is not a supersymmetric
harmonic oscillator due to the fact that $\omega _{1}\neq \omega _{2}=\frac{%
(\omega _{1})^{2}+(\alpha _{2})^{2}}{\omega _{1}}$.

From the Eq.$(5)$-Eq.$(6)$, after we perform the Bogoliubov transformation
with $\alpha _{1}=\alpha _{2}$, that preserves the algebra, Eq.$(7),$ we
obtain the harmonic oscillator with the bosonic and fermionic frequencies $%
\omega _{1}$ and $\omega _{3}$, respectively.%
\begin{equation*}
H=\omega _{1}a^{\dagger }a+\omega _{3}b^{\dagger }b
\end{equation*}%
The condition to be supersymmetric harmonic oscillator is $\omega
_{3}=\omega _{1}$, that leads the solution bellow
\begin{eqnarray*}
\omega _{1\pm } &=&\frac{1}{2}(\omega _{2}\pm (\omega _{2}^{2}-4\alpha
_{2}^{2})^{1/2}); \\ 
\omega _{2} &=&2\alpha _{2}+\xi 
\end{eqnarray*}%
$\xi$ parametrize $\omega_{2}$ at phase space.
\section{Supersymmetry Breaking at Finite Temperature for the supersymmetric Oscillator with interaction}

Supersymmetry will be broken if the thermal vacuum $|0(\beta)\rangle,$ from the supersymmetric oscillator with interaction Eq.(\ref{General model}) has non vanishing energy for a positive temperature. To introduce the temperature, we double the Hilbert space following the algorithm of TFD. Which will allow us to calculate the thermal vacuum and then the statistical average of the Hamiltonian operator of the supersymmetric oscillator with interaction, 
\begin{eqnarray}
H=F(a^{\dagger },b,b^{\dagger })a+G(a^{\dagger },a,b^{\dagger })b
\label{susy oscil with interaction}
\end{eqnarray}%
where: $F=\omega _{1}a^{\dagger }$ and $G=\omega _{2}b^{\dagger }+\alpha_{2}a^{\dagger }b^{\dagger }-\alpha _{2}ab^{\dagger }$, and $ 
\omega _{2}=\frac{(\omega _{1})^{2}+(\alpha _{2})^{2}}{\omega _{1}}$ ; $\omega_{1}$ and $\alpha _{2}$ are real parameters. With the algebra 
\begin{eqnarray}
a^{\dagger }\wedge a=-1/2;\text{\ \ \ \ \ \ \ \ \ } a^{\dagger }\bullet a=n_{b}+1/2;\text{\  }\nonumber \\
b^{\dagger }\bullet b=1/2;\text{\ \ \ \ \ \ \ \ \ } b^{\dagger }\wedge b=n_{f}-1/2; \text{\ } 
\label{algebra}
\end{eqnarray}%
$n_{b}$ $\in 
\mathbb{N} $ and $n_{f}\in \{0,1\}$. That will produce the energy.

From the Hamiltonian Eq.(\ref{susy oscil with interaction}) and using thermal field dynamics, temperature is introduced in the system, doubling the boson and fermion creation and annihilation operators. The statistical average of the Hamiltonian is given by its vacuum expectation value in the thermal vacuum. The tilde operators are defined in a similar way as the tilde Hamiltonian bellow, where $\widehat{H}$ is the generator of time translation:

\begin{eqnarray*}
\widehat{H}=H-\widetilde{H}=F(a^{\dagger },b,b^{\dagger })a+G(a^{\dagger
},a,b^{\dagger })b \\   -(F(\widetilde{a}^{\dagger },\widetilde{b},\widetilde{b}%
^{\dagger })\widetilde{a} +G(\widetilde{a}^{\dagger },\widetilde{a},%
\widetilde{b}^{\dagger })\widetilde{b}).
\end{eqnarray*}

But an elegant way to obtain the results is to use Eq.(\ref{General model}),
and the Bogoliubov transformation that preserves the algebra given in Eq.(\ref{algebra}). The Bogoliubov transformations are

\begin{eqnarray}
a_{2} =a+\frac{\alpha _{2}}{\omega _{1}}b^{\dagger }b;\text{\ \ \ \ \ \ \ \ \ \ \ \ \ \ }
a_{2}^{\dagger }=a^{\dagger }+\frac{\alpha _{2}}{\omega _{1}}b^{\dagger }b;%
\text{ } \label{bogoliubov transf} \text{\ \ \ \ \ \ \ \ \ \ } \\
b_{2} =(\exp [\frac{\alpha _{2}}{\omega _{1}}(a^{\dagger }-a)])b;\text{\ \ \ }
b_{2}^{\dagger }=b^{\dagger }(\exp [\frac{\alpha _{2}}{\omega _{1}}%
(a-a^{\dagger })]).\nonumber
\end{eqnarray}

\bigskip Transforming the supersymmetric oscillator, Eq.(\ref{susy oscil with interaction}), using the Bogoliubov transformation,
Eq.(\ref{bogoliubov transf}), we get
\begin{equation}
H=\omega _{1}a_{2}^{\dagger }a_{2}+\omega _{1}b_{2}^{\dagger }b_{2},
\label{susy harm oscil}
\end{equation}%
the algebra $a_{2}^{\dagger }\wedge a_{2}=-1/2;\ \ \ a_{2}^{\dagger }\bullet
a_{2}=n_{b}+1/2;\ \ \ b_{2}^{\dagger }\bullet b_{2}=1/2;\ \ \ b_{2}^{\dagger
}\wedge b_{2}=n_{f}-1/2.$ Showing that the supersymmetric oscillator with
interaction Eq.(\ref{General model}) could be studied as a supersymmetric harmonic
oscillator with frequency $\omega _{1}$. In order to write the thermal vacuum $|0(\beta )\rangle$, we double
the Hilbert space
\begin{eqnarray*}
|0\rangle=|0;0\rangle=|0\rangle\times |0\rangle \text{\ \ \ \ \ \ \ \ \ \ \ \ \ \ \ \ \ } \\  
|n_{b},n_{f};\widetilde{n}_{b},\widetilde{%
n}_{f}\rangle=|n_{b},n_{f}\rangle\times |\widetilde{n}_{b},\widetilde{n}_{f}\rangle.
\end{eqnarray*}

\bigskip Any operator $A(\beta )$ and the thermal vacuum $|0(\beta )\rangle$ at a
finite temperature are obtained from the zero temperature respectively by 
\begin{equation*}
A(\beta )=e^{-iG}A(0)e^{iG}\ ,\ \ |0(\beta )\rangle=e^{-iG}|0\rangle,
\end{equation*}%
where $G=-i\theta (\beta )(\widetilde{b}b-b^{\dagger }\widetilde{b}^{\dagger
})-i\theta (\beta )(\widetilde{a}a-a^{\dagger }\widetilde{a}^{\dagger })$, $%
\beta =1/\kappa T$, $\kappa$ is the Boltzmann constant with $\tan \theta
(\beta )=e^{-\beta \omega _{1}/2}$.

\begin{figure}[htb]
\includegraphics[scale=1.0]{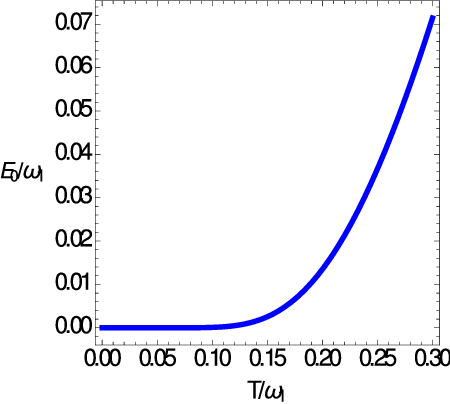}
\caption{Plot of the vacuum energy ${E_{0}}/{\omega _{1}}$, as function of ${T}/{\omega _{1}}$.}
\label{fig1}
\end{figure}

The thermal energy of the themal vacuum is given by
\begin{equation*}
E_{0}(\beta )=\langle 0(\beta )|\omega _{1}a_{2}^{\dagger
}a_{2}+\omega _{1}b_{2}^{\dagger }b_{2}|0(\beta )\rangle=
\end{equation*}%
\begin{equation}
\omega_{1}\left( \frac{e^{-\beta \omega _{1}}}{1-e^{-\beta \omega _{1}}}+%
\frac{e^{-\beta \omega _{1}}}{1+e^{-\beta \omega _{1}}}\right) .
\label{Energia no vacuo termico}
\end{equation}%
This shows that the supersymmetric is broken at $T > 0$, Fig \ref{fig1}.

The Witten index is:%
\begin{equation*}
\Delta (\beta )=\frac{1-e^{-\beta \omega _{1}}}{1+e^{-\beta \omega _{1}}}.
\end{equation*}%
The action of the supersymmetric charges over the thermal vacuum is given by:
\begin{eqnarray}
G_{2S}|0(\beta )\rangle = a_{2}^{\dagger }b_{2}|0(\beta )\rangle= \text{\ \ \ \ \ \ \ \ \ \ \ \ \ } \\ \nonumber \frac{e^{-\beta \omega
_{1}/2}}{\left[ (1-e^{-\beta \omega _{1}})(1+e^{-\beta \omega _{1}})\right]
^{1/2}}|\chi _{1}(\beta )\rangle, \\
\overline{G_{2S}}|0(\beta )\rangle= b_{2}^{\dagger }a_{2}|0(\beta )\rangle= \text{\ \ \ \ \ \ \ \ \ \ \ \  } \\  \nonumber \frac{e^{-\beta \omega _{1}/2}}{\left[ (1-e^{-\beta \omega _{1}})(1+e^{-\beta
\omega _{1}})\right] ^{1/2}}|\chi _{2}(\beta )\rangle,
\end{eqnarray}%
where $|\chi _{1}(\beta )\rangle$ and $|\chi _{2}(\beta )\rangle$ are the Goldstino
states at finite temperature, that are produced from the vacuum by applying
supersymmetric charges.

\section{Conclusion}

An oscillator model with polynomial interactions is analysed and we show
conditions that lead the same to a supersymmetric harmonic oscillator with
interaction. The temperature is introduced in this oscillator model with interaction through TFD formalism. Reducing the model to a supersymmetric
harmonic oscillator, we analyze the supersymmetry breaking at finite temperature. The fermionic and bosonic quantum corrections in a supersymmetric
theory tend to cancel. The thermal effects are additive and the thermal
energies are positive. The statistical average of the\ Hamiltonian at finite temperature is not zero.

At $T=0$ from the same parameter $\alpha _{2}$ of polynomial interaction and frequency $\omega _{2}>2\alpha $ we obtain two bosonic frequencies $\omega
_{1}$ that gives supersymmetric harmonic oscillator ($\omega _{1-},\omega
_{1-}$) and ($\omega _{1+},\omega _{1+}$). For the case $\omega
_{2}=2\alpha _{2},$ we have only one supersymmetric harmonic oscillator for the
same polynomial parameter of interaction, since $\omega _{1-}=\omega _{1+}.$

An upcoming and possible application can be implemented in the work Searching Neutrino-Nucleus interaction in M\"ossbauer Spectroscopy \cite{Neutrino}; The break of spherical symmetry perceived by the wave function of the valence neutron field, in some conditions (Woods-Saxon potential), could be thought as a bosonic harmonic oscillator with $\omega_{1}$ frequency. 
The vibrations between two modes could be described here as a two level system of a narrow band of vibration.  That vibration around the $3/2$ and $5/2$ spin states of the $Fe$ nucleus could be thought as a fermionic harmonic oscillator with $\omega_{2}$ frequency. If $\omega_{1}$=$\omega_{2}$ we have the supersymmetric case. 
\vspace{-\baselineskip} 

\section*{ACKNOWLEDGEMENTS}

Professor Gilmar De Souza Dias is grateful to the University of Alberta,
Canada for the office and other facilities during all his stay in the UAlberta, when interesting works were made and interesting ideas were developed.This work being an example.

\end{document}